# Andreev Reflection Spectroscopy in Transition Metal Oxides

Karen A Yates and Lesley F Cohen

*Physics Department, The Blackett Laboratory, Imperial College London, UK, SW7 2AZ*

Here we review the literature concerning measurement of the Andreev reflection between a superconductor (S) and ferromagnet (F), with particular attention to the case where the ferromagnet is a transition metal oxide. We discuss the practicality of utilisation of the current models for determination of the transport current spin polarisation and examine the evidence for Andreev bound states.

I: Introduction

Point contact Andreev spectroscopy (PCAR) has been widely used to measure the transport spin polarisation of conductors since the seminal 1998 paper by Soulen et al., [1]. The transport spin polarisation, *P*, is the degree to which the carriers of electrical current possess an excess of one direction of spin. The use of PCAR to determine *P*, is based on the premise that at voltages below the superconducting gap voltage and at an interface between an s-wave superconductor and a normal metal, an electron with one spin direction, (eg. spin up), couples with an electron with the opposite spin (ie. spin down) passing into the superconductor as a Cooper pair whilst requiring the retro-reflection of a hole of the opposite spin (ie. spin down) back into the normal metal. This leads to a doubling of the conductance at voltages below the gap voltage, for a review see [2,3]. As the transport spin polarisation of the metal increases, the probability of retro-reflection for the hole is reduced and consequently the Andreev process is suppressed. This leads to a reduction in the conductance at voltages below the superconducting gap voltage. The degree to which the Andreev signature (i.e. the doubling of the differential conductance at bias voltages less than the superconducting energy gap) is suppressed is an indication of the degree of *P* [1]. In the years shortly following 1998, the interest in using PCAR to determine the degree of transport spin polarisation *in* materials for spintronic applications increased with time [4,5]. The original theoretical description of Andreev conductance across a superconductor-normal metal (SN) contact, the "BTK" model [6] was soon adapted to incorporate the parameter of transport spin [7]. The ease with which the Mazin-BTK (M-BTK) method could be implemented, both experimentally and computationally resulted in a plethora of papers investigating *P* for various different materials [4,8,9,10,11,12,13,14]. As more measurements became available however, many papers quickly demonstrated that a detailed description of the interface between a superconductor and a ferromagnet was lacking [15, 16, 17], leaving some spectroscopic features poorly understood and outside the scope of the original model.

One such aspect of complexity was the description of finite scattering at the interface, which also acts to suppress the Andreev process [18]. In the original BTK model this scattering was accounted for using a delta function potential Z [6, 18]. Much experimental data showed a dependence of the extracted spin polarisation *P*, on this interfacial scattering parameter, Z [12,19]. This dependence lay beyond the original Mazin-modified BTK (M-BTK) model [12, 20]. Similarly it was shown that superconducting point contacts onto F materials often showed spectra with enhanced sub-gap conductance together with a suppressed order parameter [5,21], requiring the introduction into the MBTK model of a non-thermal spectral broadening parameter $\Gamma$ in order to describe the experimental data [5, 11,22]. It was suggested that this broadening could result from real physical phenomena occurring in the contact [5,17,22] but could also be the result from measurement artifacts [23], or errors introduced in the application of the MBTK model to the data [17].

At a similar time, theoretical work to understand the order parameter symmetry of the (at that time) recently discovered cuprate superconductors emphasised that under certain conditions sub-gap states and zero bias anomalies could occur in ordinary SN contacts [24]. Similarly, interference effects occurring when the N (or F) material was a thin film of a specific thickness, could cause Andreev Bound States (ABS) to appear at finite energy within the conductance spectra [3,25]. As most of the materials studied were neither thin films, nor materials containing inhomogeneous magnetic scattering layers [24], in this early work, the existence of an Andreev Bound State or states was not explicitly taken into account. However, concomitantly, the nascent field of S/F/S (so called $\pi$ junctions) was beginning to establish detailed understanding of the influence of the ferromagnetic exchange field on the superconducting phase difference across the Josephson junction [26, 27, 28, 29, 30,31] including the possibility of the formation of a different type of Andreev Bound State in the junction itself [27,29,31,32]. The concern expressed at this time was that the treatment of the transparency of the junction based merely on the probability of the occurrence of the Andreev process, may be incomplete and therefore the extracted polarisation using these assumptions might be subject to significant error. Indeed, more recent models of Andreev reflection across an SF interface, evolving to some extent from the $\pi$ junction work, proposed the ABS as an intrinsic component to the conductance across any constriction SF contact [20,33].

This review summarises the literature regarding point contact studies of SF structures focussing on the transition metal oxides such as $CrO_2$ in which very high levels of polarisation have been observed [8,13,14,21]. The varying predictions (with temperature or magnetic field) for ABS will be reviewed in order to elaborate the conditions under which Andreev spectroscopy can be employed to probe transport spin polarisation.

In the first part of this paper (part II), the early difficulties of Andreev reflection as a probe of spin polarisation will be considered. These `traditional' difficulties arose as a consequence of the direct application of the Blonder-Tinkham-Klapwijk (BTK) model to SF interfaces. No account was taken either in theory or in practise in these models for spin specific transport phenomena. Mechanisms producing ABS will be considered here as well as alternative explanations for higher than expected sub-gap conductance. In part III, models of conduction in which spin is a parameter of the model are examined. In such models, ABS arise as a consequence of the spin transport and the effects of the ABS on the measured Andreev reflection spectra are considered. The question is explored as to whether a unique value of transport spin polarisation can be obtained within such a model due to the number of parameters involved. Part IV considers the experimental evidence for ABS while part

V considers the experimental measurements of SF interfaces in general. The final part of the paper suggests some routes forward so as to make Andreev spectroscopy useful in the toolkit of techniques available for the measurement of spin polarisation.

II: Traditional obstacles to the determination of P

The celebrated success of the BTK model was the incorporation of the concept of a dirty interface using the delta function parameter Z. Z [6] reduces the transparency of the Andreev contact [18]. For low Z the barrier is transparent to conduction electrons and conduction into the superconductor at sub-gap bias occurs via Andreev reflection. At high Z, the contact is tunnel-like. For intermediate values of Z (ie. all realistic point contacts) conduction is modelled as involving contributions from both Andreev reflection and tunnelling currents. The BTK model gave the conductance as a function of Z for unpolarised contacts [6,18]. The M-BTK model incorporated both *P* and Z using the Landau Buttiker formulism. The focus of early questions once the M-BTK model was established, related to the meaning and nature of Z and the physics that might be contained within it[12,19,22]. As both *P* and Z act to suppress the conductance and as, ultimately the `measurement' of *P* via Andreev spectroscopy involves *fitting* a conductance curve, it was recognised that the effects of Z and *P* could become indistinguishable, particularly in the presence of thermal (and non-thermal) smearing [9,11]. This P-Z compensation effect however did not initially prove a problem as the community quickly developed methods of fitting the data so as to reduce the P-Z ambiguity [11,12]. Nonetheless, further complicating factors were identified: was there one 'Z' for the contact, or did each separate spin band have a different effective energy barrier to overcome [15]? Was it even valid to assume a 1D type transport for a 3D system [17]?

Although Z had initially been introduced merely to account for scattering, it was shown to have additional physical meaning [2,22]. Even the effect of Fermi velocity mismatch at the SF contact provided a contribution to Z [17,18,21] which in principle, results in a lower bound for Z ($Z_{eff}$) given by [18]:

$$Z_{eff} = \sqrt{Z^2 + \frac{(1-r)^2}{4r}}$$

Where *r* is the ratio of the Fermi velocities of the two materials forming the contact and Z is any additional scattering [18]. Naturally this minimum Z would be different for the two different spin bands [17]. However, even assuming an average effective mismatch, the fit to an experimental data-set that yields Z = 0 is unphysical. The Z=0 problem is a feature present in the PCAR results of many highly polarised materials [34], particularly the fully spin polarised $CrO_2$ [21]. As well as *P* and Z, other parameters were introduced into the fit routine including the superconducting gap parameter, $\Delta$, and a spectral smearing or broadening parameter, $\omega$ (or $\Gamma$) [11] (or equivalently either an effective temperature [5,9] or spreading resistance [12]). The physical justification given for the introduction of these parameters was for example, that with a point contact under mechanical stress, $\Delta$ may not be the bulk value [11] and $\omega$ can include many factors of broadening [22,23]. Over many contacts *P* often appeared to have a dependence on Z [12,19,34,35], figure 1a. The qualitative (though not the quantitative) form of this PZ relation was independent of whether the contact was fitted in the limit of ballistic (small contact size) or diffusive transport (contact diameter comparable or larger than the mean free path of the studied material) (fig 1a) [12]. Various theories were proposed for the origin of this dependence [19], although no unique mechanism or explanation has been identified [12]. Nonetheless, contacts showing high polarisation repeatedly have been observed to have (unphysically) low values of Z and the method of extrapolation to zero Z as a determinant of bulk spin polarisation became quite popular [34,35].

These two problems are both present in Andreev spectroscopy measurements of the highly spin polarised material $CrO_2$, figs 1a, 2. For this 100% spin polarised material and transition metal oxide, the highest polarisation has often been determined from spectra that when fitted, resulted in very low or zero Z values [8,20,21,34,36]. A zero Z means that there is no interface scattering and the suppression of the sub-gap conductance is entirely due to finite transport spin polarisation. In the context of this review on the potential influence of ABS on the conductance spectra of SF contacts, this becomes an interesting observation: The effect of ABS is to increase the sub-gap conductance without any influence on the conductance at the gap voltage (i.e. contribution to the coherence peaks associated with finite Z). If such a spectrum with ABS present were to be fitted without accounting for these ABS, the extracted parameters would either predict an artificially lower *P* or Z and/or include a significantly enhanced non-thermal broadening term to explain the finite states in the gap [5]. It is interesting to speculate that the so called 'Z=0 problem' is actually an anomaly in the fitting due to the presence of ABS.

For the particular case of $CrO_2$ interrogated with a Pb contact, the Fermi velocity mismatch results in a minimum Z of ~ 0.26 [21]. As figure 2 shows, those contacts with an apparent extracted *P* > 85%, all have an extracted Z parameter lying below this minimum. This trend also appears in data taken from the literature for other Andreev contacts onto $CrO_2$ as shown in the summary figure 2d [20 and references therein]. In addition to the PZ dependence, plotting the other extracted parameters against Z reveals that contacts showing a high polarisation are often associated with a fitted suppressed superconducting gap parameter value [21], figure 2b, 2e. Although significant non-thermal broadening also exists in the spectra shown in figure 2c, there is no discernable correlation between the degree of suppression of the gap feature and the magnitude of the spectral broadening, from which one can rule out inelastic scattering as a cause of gap suppression [21]. Data taken from the literature fitted using a series resistance rather than a smearing parameter [20], showed a similar trend for $CrO_2$ (figure 2f). Similar results showing depressed superconducting gap values associated with spectra showing enhanced spectral broadening were observed in (Ga,Mn)As [5].

It is important to note that there are certain artefacts that could cause such an enhancement of the sub-gap conductance. As shown by Zhu et al., [24], the presence of a magnetic scattering layer at some distance within the normal metal from the SN interface, produces sub-gap states caused by multiple carrier reflections between the superconductor, the normal metal and the magnetic scattering layer. Similarly, interference effects causing sub-gap Andreev Bound States (ABS) are expected in simple SF interfaces if the ferromagnet is a thin layer [3,25]. Zero bias anomalies (which in the case of significant non-thermal broadening could appear as an enhanced sub-gap conductance) can also be caused by Joule heating in the point contact area if the contact is not in the ballistic limit (where the ballistic limit is defined as the contact diameter less than both the mean free path in the ferromagnet and the superconducting coherence length [2,18]). However, although these physical effects cause states in the gap, they are all very specific phenomena which can be avoided by judicial choice of sample and so should not be the cause of enhanced sub-gap conductance observed systematically in $CrO_2$ [21] and (Ga,Mn)As [5].

Other predicted phenomena such as a resonant enhanced proximity effect which can lead to both zero bias and finite bias peaks [26], or magnon assisted sub-gap transport [37] are more difficult to avoid and therefore cannot be categorically excluded as an explanation of the behaviour in figure 2. However, despite reports of enhanced sub-gap conductance due to the proximity effect in superconductor-semiconductor structures [38,39], the effect of these predicted proximity issues in SF structures was assumed to be within experimental error (which is considerable for the point contact technique [11]), or controllable if the contact size was kept in the true ballistic limit [3,15,17]. Indeed, measurements of *P* via the Andreev technique for moderately spin polarised materials without accounting for these anomalous effects, give results consistent with those of other techniques [11,17]. The puzzle of the enhanced inelastic scattering (and reduced gap value) in contacts fitted using the M-BTK theory onto (Ga,Mn)As [5], $CrO_2$ [21] and Pt [17] therefore remained.

III: Spin Dependent Scattering

The first models that considered the effect of the spin of the particle as it traversed the interface did so in terms of the spin transparency of the SF contact [15,17,40], introducing different transmission coefficients for the different spins. The differences in the polarization values determined by fitting to this type of model compared with M-BTK, was reported by [41]. Although the effect of the barrier transparency on the different spins was considered in these models, the possible effect of a phase difference between the wavefunctions of the different spins was not. Consequently, although there were differences in sub-gap conductance values between the spin-transparency models and the M-BTK, at this stage Andreev Bound States were not discussed [41].

ABS were included in the models that considered that the ferromagnet produced a phase difference between the two spins as well as a spin dependent transparency difference. These models extended the understanding of the phase difference between the spins in superconducting $\pi$ junctions [27,31] to simple, single interface SF point contacts [20,28,33,42,43,44]. The models introduced a spin mixing angle, $\theta$, (an interfacial property that results in the different spins having a different phase as they cross the SF interface). This spin mixing angle can significantly affect the sub-gap conductance and in particular, should result in the appearance of Andreev Bound States in the spectra [33,42]. The spin mixing angle also acts to lower the intensity of the coherence peaks at the gap voltage and shift the spectral weight of this contribution to sub-gap voltages via the ABS [45]. An effect which would appear, if modelled by the M-BTK model as a reduced Z, and high non-thermal broadening.

The differences between the M-BTK and the spin mixing model (SMM) as described in ref [20,33] can be illustrated by considering the limiting case of 100% spin polarisation, such as expected for $CrO_2$ [20]. For the M-BTK model, the absence of one spin species leads directly to a complete suppression of $G_0$. A finite conductance at zero bias for a 100% spin polarised material can only be achieved, with the M-BTK model, in the presence of finite broadening of the quasiparticle density of states, such as exists at finite temperature or through inelastic scattering. It is the sum of these broadening parameters that determines the temperature evolution of $G_0$ [22] and particularly it is these features that would be responsible for the `flattening off' of the $G_0(T)$ at low temperatures within the M-BTK model [20]. The case is entirely different within the SMM. Within this model, each spin picks up a phase as it traverses the SF boundary. This phase difference allows fully spin polarised materials to nonetheless form a Cooper pair of the triplet form ($|\uparrow\downarrow> + |\downarrow\uparrow>$). It therefore becomes possible for Cooper pairs from S to travel to F that is, the Andreev process is not completely suppressed and the sub-gap conductance is no longer, necessarily zero. ABS form at finite voltages, the value of which is determined by the spin mixing angle, $\theta$ by [31,33]:

$$\epsilon_{pole} = \pm\Delta\cos\left(\frac{\theta}{2}\right)$$

At finite temperature, the ABS features have a finite voltage width and so contribute to $G_0$. As the temperature is reduced (the thermal broadening decreases), the ABS become more clearly defined in principle and the zero bias conductance reduces to its highly polarised value of zero. Although the special case of $\theta = \pi$ would result in a bound state at zero bias, the geometry of the contact, barrier and the Fermi surfaces of S and F combine to make such high values of $\theta$ unlikely [33]. Beri et al., [43] also considered the case of point contacts onto half metals, taking spin mixing into account explicitly, and also predicted $G_0 = 0$ at T=0 K.

In these models therefore, excess sub-gap conductance is observed owing to differences in spin transport across the contact and the generation of the spin triplet proximity effect in the ferromagnet [20,28,43]. If such effects are present in experimental spectra and fitted according to a model that does not take them into account (for example using the M-BTK model), the resultant fitting would give inaccurate values for all fitting parameters, but of most concern in this context, of *P*. Indeed, data taken on a material showing an intermediate spin polarization, (Ga,Mn)As, required substantial inelastic scattering (as well as a much reduced gap value) to be fitted within the MBTK model [5]. Data taken on (Ga,Mn)As but fitted using SMM could not completely accommodate the anomalously low values of $\Delta$ although the apparent high smearing was a natural consequence of the ABS [45].

In order to develop these ideas further, Löfwander et al., [20] have suggested two key tests of the M-BTK model versus the SMM and also, a method of refining the fitting parameters. Firstly, in principle, merely measuring the zero bias conductance of an SF contact as a function of temperature may indicate whether the MBTK or the SMM is the more appropriate model [20]. The temperature dependence of the zero bias conductance, $G_0$, (normalised to the zero bias conductance just above $T_c$), shows different behaviour for the M-BTK model compared to the SMM (figure 3a,b). On the positive side, this difference can be observed without any fitting of the data; it is a direct measurement. On the negative side, the low temperatures needed in practise to settle the question definitively and the additional broadening encountered in any real-world experiments [22], combine to make use of this technique (particularly if using a mechanical point contact technique) difficult [46]. Measurement of the excess current of the contact acts as a complementary method to ensure that the extracted fitting parameter, Z, is consistent with the model used [20] (fig 3c,d). For example, in figure 3c, the excess current (normalised to the normal conductance) for the M-BTK model as a function of Z and polarisation (curve) shows quite different behaviour to the excess current, for fully spin polarised materials, as a function of the fitted Z and the spin mixing angle, θ, figure 3d. Although the measurement of I(V) is implicit in the measurement of dI/dV, the fitting routine for each needs to yield the same parameters and this acts as a quasi-independent set of tests.

This combination of reduced temperature together with a cross-check of two different measurements to extract the parameters from the spectra suggests a possible route forward if the SMM is to be used in order to determine the spin polarisation from point contact Andreev reflection measurements.

IV: Experimental evidence for ABS in SF Andreev spectra

To date, there have been no definite experimental signatures of ABS caused by spin mixing in highly spin polarised Andreev point contact experiments. Partly this could be attributed to experimental constraints: To observe the ABS, high values of θ [33] are needed, with fairly low transparency [27] and at low temperatures [31,33]. When coupled with the fact that point contact spectra are assumed to be made up of many nanocontacts rather than one single SF contact [11], resulting in the likely distribution of features, it is less surprising that ABS features have yet been reported from point contact Andreev spectroscopy.

There is however one direct observation of ABS reported in tunnel spectra of Al-$Al_2O_3$-Fe junctions [47], figure 4. In total 30 contacts were measured in this study, a small number of which showed contacts with symmetric ABS, as predicted by Grein et al [33]. For these contacts, the observed magnitude of the conductance peaks was around three orders of magnitude lower than the conductance of the junction at the gap voltage, figure 4b, note scale. In order to unambiguously identify the peaks as ABS rather than, eg. minigap features, Hübler et al., measured the effect of a magnetic field on the conductance peaks. Hübler et al., argued that if the features were attributable to a minigap as a result of the proximity effect, they should disappear at high values of magnetic field. In contrast, the features that they observed changed with field as if they were subject to the Zeeman effect moving linearly with a slope of $\mu_B B$ with field, only disappearing as the magnetic field increased above the critical field of the Al superconductor [47].

Although the size of the features and the technical challenges identified by Hübler et al., make identification of ABS using a point (or planar) Andreev conduction technique difficult, the use of the magnetic field to identify ABS, together with the earlier identified use of temperature, offer a possible method for identification of features on the Andreev spectrum as ABS induced by spin mixing. Furthermore, if the conductance spectra can be shown to reveal ABS at specific voltages, equation (2) gives θ, thereby reducing the number of fitting parameters needed to fit the data within the SMM.

V: Experimental measurements within the BTK and SMM

Although the direct observation of ABS remain a challenge, nonetheless it is possible that it is ABS features that broaden at finite temperature into a continuum [27] that contributes to the sub-gap conductance in SF point contact spectra [33]. Nonetheless, even in the presence of a continuum of ABS, the behaviour of the $G_0/G_n$ curve as described by Lofwander [20] should be relevant as a test of the SMM or MBTK models. With zero non-thermal broadening, the M-BTK model predicts an exponential-like decrease of the conductance with temperature as shown by the blue dash-dotted line in figure 5. In contrast, the SMM with a spin mixing angle of $\theta = \pi/2$ shows a slower decrease of $G_0$ with temperature (black lines on figure 5). For zero non-thermal broadening therefore it should be easy to distinguish MBTK from SMM. However, a moderate amount of non-thermal broadening present in the data, causes the predictions of the two models to become harder to distinguish [46], as shown by the blue dashed line in figure 5 where a small amount of non-thermal broadening has been included into the M-BTK model. Mechanical point contacts such as those used to take data onto $CrO_2$ inevitably show a significant amount of non-thermal broadening [48]. Data taken comparing planar contacts with the mechanical point contacts of Pb onto Cu and Co showed that while the planar technique could achieve spectra in which the broadening parameter approached the thermal limit, mechanical point contacts often showed a broadening parameter significantly above this [48]. Although many details of the interface and the material studied may contribute to such non-thermal broadening, [17,22], the presence of this non-thermal broadening can introduce a non-trivial temperature dependence into the behaviour of the contact. Such a broadening parameter in the case of magnetic materials may indicate the presence of underlying ABS within the spectra [45] although it may also be due to the presence of spin flip scattering [17]. However, regardless of whether it is indeed indicative of ABS, such broadening would affect the distinctiveness of the $G_0/G_n$ test of figure 5.

The temperature dependence of a mechanical point contact using Pb as a superconductor onto a $CrO_2$ thin film grown on $TiO_2$ is shown in figure 5a. In addition to the expected suppression of the conductance at $|V|<\Delta$, there is a non-linear background that exists until $T>T_c$. In order to compare the data to the theoretical predictions of the SMM in figure 3, the zero bias conductance at any given temperature was normalised to the zero bias conductance measured at 7.3K ($T_c \sim 7.2K$). $G_0/G_n$ data extracted in this way for $CrO_2$ films grown onto $TiO_2$ (blue data) [46] and $Al_2O_3$ (dark yellow data) [21] are shown in figure 5b. Also shown are $G_0/G_n$ data extracted from spectra in the literature for $CrO_2$ grown onto $TiO_2$ (blue data) [14] and the equally highly spin polarized $La_{0.7}Sr_{0.3}MnO_3$ [4]. In the case of the contacts onto $CrO_2$, although none of the $G_0/Gn$ (T) curves fall on that expected for the M-BTK limit of 100% polarisation, as discussed above once additional non-thermal broadening is included, the predicted M-BTK model curves approach those of the experimental data at temperatures far above 0K [46]. The data is therefore preliminary and neither confirms nor refutes one or other of the models. The data from the literature on LCMO by contrast is taken to lower temperature [4] and appears inconsistent with the M-BTK prediction, however non-thermal broadening mechanisms are also likely to be contributing in these junctions and only further investigation taken to lower temperatures could help definitively establish the correct behaviour.

In summary, although ABS have been predicted as an intrinsic component of SF point contacts, there has only been one direct observation of such an ABS [47]. Partly this can be attributable to the fact that it is difficult to make one, well defined contact that satisfies the conditions needed to observe clearly defined ABS features [33]. Nonetheless, their presence in the spectra should be revealed by the temperature dependence of the zero bias conductance which will approach zero at 0K for a half metallic contact [20]. Measurements of the temperature dependence of $G_0(T)$ however have so far proved inconclusive as it is necessary to measure the contacts down to very low temperature and, probably to use planar contacts to do so. The presence of ABS in the spectra would result in reported data of low polarisation or significant additional non-thermal smearing if fitted within the M-BTK models. This necessitates the fitting of the data with models which properly account for the potential formation of ABS. Nonetheless, such models introduce more fitting parameters than the M-BTK, making practical determination of the spin polarisation less tractable. If however, future point contact measurements are made to lower temperature, recording both dI/dV and IV (in sufficient detail that $d^2I/dV^2$ can also be determined), it should be possible to determine the spin mixing angle, $\theta$, if ABS are observed clearly and to cross check the remaining fit parameters with the dI/dV and IV measurements outlined in ref [20]. These measurements can also check the applicability of the MBTK and SMM models under these circumstances. Although this reduces the "quick and easy" advantage of Andreev reflection as a method to determine the transport spin polarization of a ferromagnetic metal, it potentially enhances the value of the method in terms of the rich knowledge about the intimate interactions at SF interfaces it reveals.

Figures:

Figure 1:

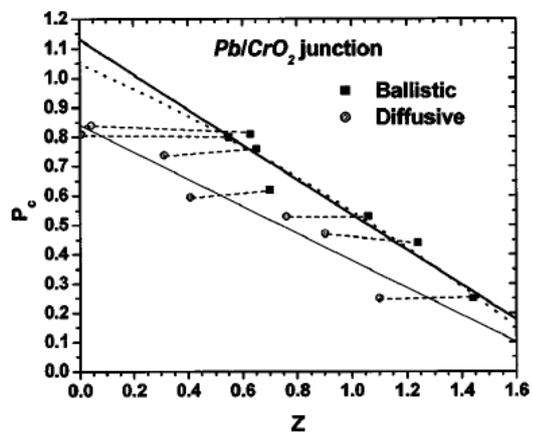

Figure 1, (a) PZ relation for the ballistic and diffusive limit fitting for a Pb contact onto a $CrO_2$ (film). Reprinted figure with permission from [12], copyright (2004) by the American Physical Society.

Figure 2:

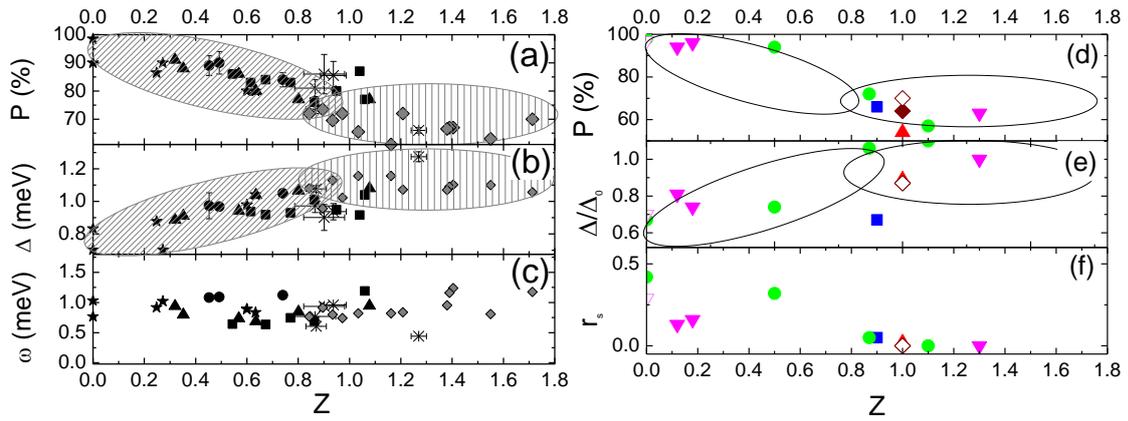

Figure 2: Parameters as a function of Z extracted via the MBTK model for Pb contacts onto $CrO_2$ thin films, Reprinted with permission from [21]. Copyright [2007], AIP Publishing LLC (a) Polarisation, (b) $\Delta$ (c) broadening and from the literature ref [20] (d) P (e) $\Delta/\Delta_0$ where $\Delta_0$ is the expected gap of the superconductor and (f) the spreading resistance $r_s$. Symbols are for the data from (■) [1], (▲) [49], (●) [13],(▼) [8], (∇) [36], (◆) [50], (◇) [51].

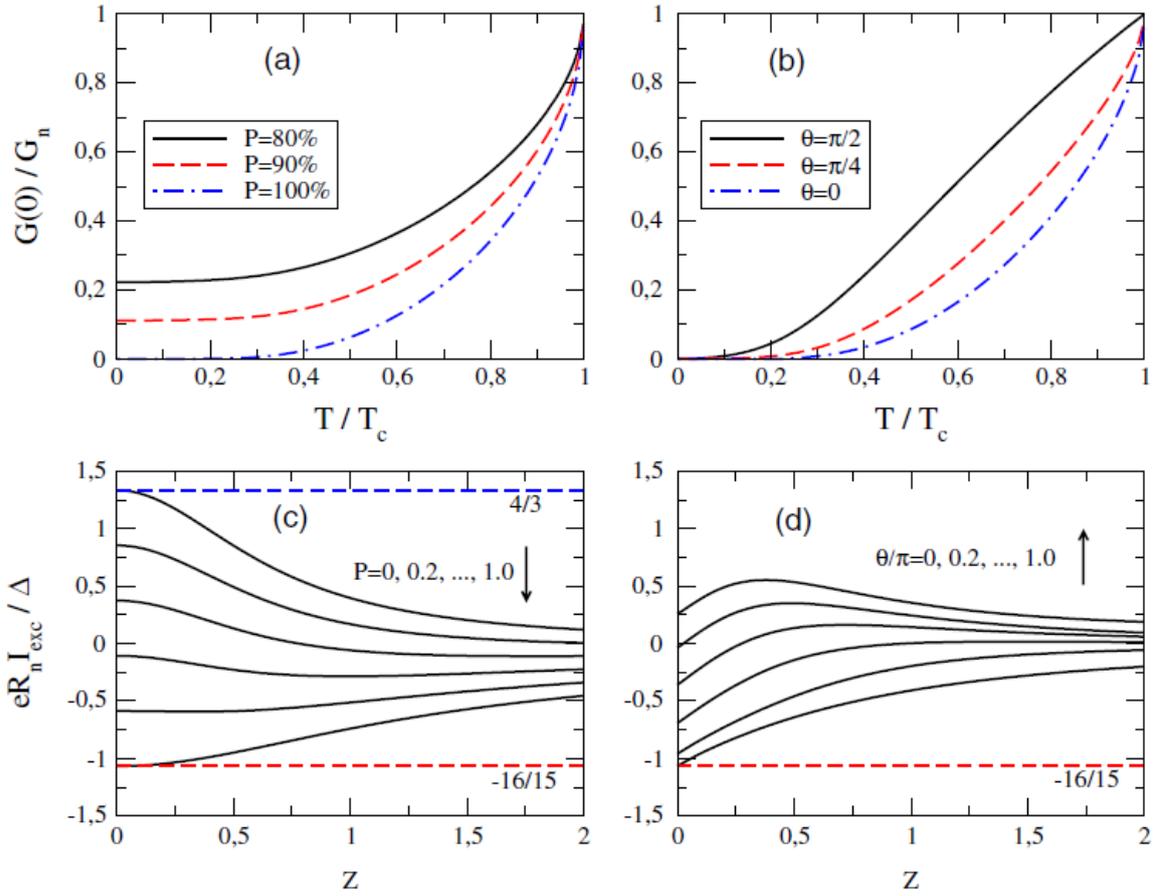

Figure 3: (a) $G_0/G_n$ in the MBTK model and (b) the SMM for P = 100% as a function of spin mixing angle, $\theta$. (c) the excess current in the MBTK model as a function of Z, (d) in the SMM. Reprinted figure with permission from [20]. Copyright (2010) by the American Physical Society.

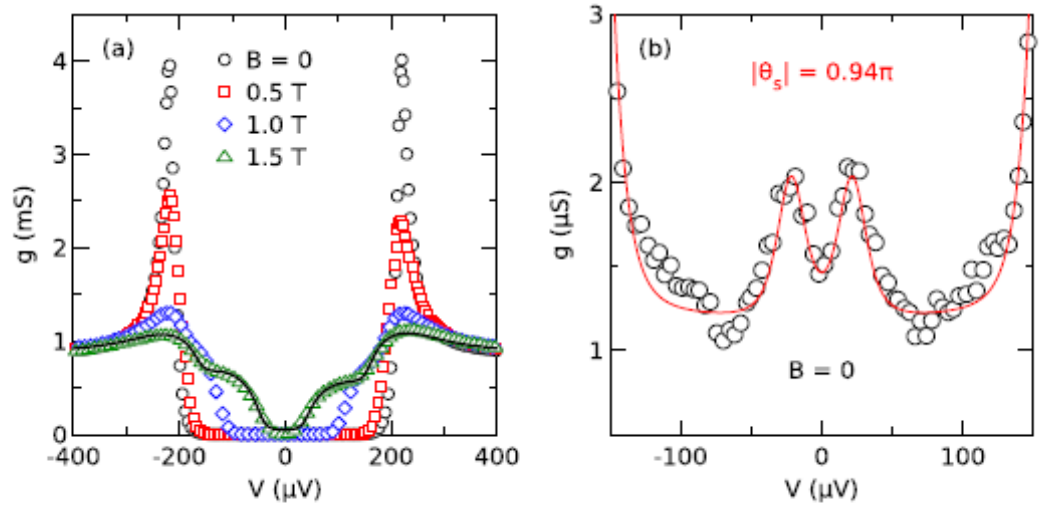

Figure 4: Observation of Andreev Bound States in tunnelling spectra between Fe and Al. Note the scale difference between figures (a) and (b). Reprinted with permission from [47]. Copyright (2012) by the American Physical Society.

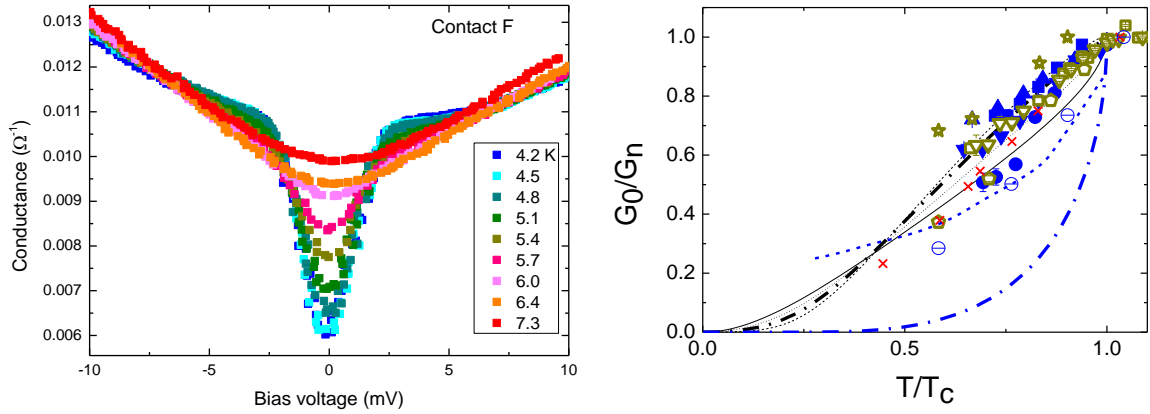

Figure 5: Temperature dependence of a Pb contact onto a CrO$_2$ thin film, adapted from [46] (b) extracted $G_0/G_n$ data for CrO$_2$ films onto TiO$_2$ (blue symbols, [46],ϴ [14]), Al$_2$O$_3$ (Dark yellow, [21]), and LCMO (red, x) taken from Nadgorny et al. [4]. Black lines are theoretical predictions of the SMM model for Z = 0.1, 0.26, 0.5 and 1.0 and θ = 0.5. Blue dash-dotted line is the MBTK for P = 100%. Blue dashed line represents the expectation of the MBTK P=100% with 0.7meV additional non-thermal broadening added to the temperature dependence.